\documentclass[final,1pt,times]{elsarticle}

\usepackage[a4paper,margin=1.8cm]{geometry}
\usepackage{amssymb}
\usepackage{amsfonts}
\usepackage{amsmath}
\usepackage{graphicx}
\usepackage{bm}
\usepackage{braket}
\usepackage{epstopdf}
\usepackage{xcolor}
\usepackage[colorlinks=true,linkcolor=blue,citecolor=blue,urlcolor=blue]{hyperref}
\usepackage{makeidx}
\usepackage{mathrsfs}
\usepackage{dcolumn}
\usepackage[center]{subfigure}
\usepackage{color}
\usepackage{indentfirst}
\usepackage{array}
\usepackage{booktabs}
\usepackage{float}
\usepackage{multirow}
\biboptions{sort&compress}

\usepackage{amssymb}

\usepackage{amsmath}

\journal{Chaos, Solitons \& Fractals}
\begin{document}

\begin{frontmatter}

\title{Elastic Modulus in One-Dimensional Quantum Droplets} 

\author[a]{Rui Zhang\fnref{equal}}
\author[a]{Tianmiao Zhang\fnref{equal}}
\fntext[equal]{These authors contributed equally to this work.}
\author[a]{Huan-bo Luo\corref{cor1}}
\author[a]{Zibin Zhao\corref{cor2}}

\cortext[cor1]{Corresponding author: huanboluo@fosu.edu.cn}
\cortext[cor2]{Corresponding author: zzbphys@gmail.com}

\affiliation[a]{organization={School of Physics and Optoelectronic Engineering, Foshan University},
	city={Foshan},
	postcode={528000},
	country={China}}



	


\begin{abstract}
	
Quantum droplets (QDs) are self-bound states of ultradilute quantum fluids stabilized by the interplay between the Lee-Huang-Yang (LHY) quantum-fluctuation correction and 
the mean-field interaction, providing a useful platform for exploring macroscopic quantum phenomena. Recent studies on three-dimensional QDs have introduced the concept of 
bulk modulus and revealed its connection with the breathing-mode frequency, thereby linking the elastic response of QDs to their collective dynamics. Motivated by this progress, 
we investigate the elastic modulus of one-dimensional QDs. Based on a super-Gaussian variational ansatz, we systematically derive the elastic modulus $B$ and analyze its dependence 
on the interaction strength and particle number. The analytical predictions are further validated by numerical simulations based on imaginary-time evolution and the spatial scaling 
method. We also establish a quantitative relation between the elastic modulus and the eigenfrequency $\Omega$ of the breathing mode. In addition, by incorporating corrections to 
the droplet width beyond the Thomas-Fermi approximation, we obtain the dependence of the ratio $\eta=B/\Omega^2$ on the control parameters $g$ and $N$. Unlike the three-dimensional 
case, where the corresponding ratio follows a simple power-law scaling, the one-dimensional system is affected by the soliton-to-droplet crossover, leading to a more intricate 
dependence of \(\eta\) on \(g\) and \(N\). Our results show that, in 
the high-particle-number regime, the elastic modulus asymptotically approaches a limiting value determined mainly by the interaction strength, whereas in the low-particle-number 
regime it depends on both the particle number and the interaction strength. These findings provide a theoretical basis for characterizing the macroscopic elastic properties of 
one-dimensional QDs and for connecting their mechanical response with collective excitation dynamics.

\end{abstract}

\begin{keyword}

Quantum droplets\sep
variational method \sep
elastic modulus \sep
breathing-mode frequency 

\end{keyword}

\end{frontmatter}


\section{Introduction}
Elastic modulus is one of the fundamental quantities characterizing the mechanical response of matter. In classical continuum mechanics, elastic moduli measure 
the resistance of a material to external deformation, while in fluids the relevant quantity is usually the bulk modulus, which describes the resistance of a 
fluid to compression \cite{Elasticity,Continuum}. The bulk modulus is closely connected to the equation of state and compressibility, and it also characterizes the response of a fluid to 
compressional oscillations. Therefore, the concept of modulus provides a natural link between the static mechanical properties of a system and its dynamical response.

Quantum droplets (QDs) are self-bound quantum fluids stabilized by the balance between mean-field interactions and beyond-mean-field quantum fluctuations \cite{LHY-3d,LHY-1D}. Since their 
theoretical prediction and experimental realization in dipolar Bose gases and binary Bose mixtures \cite{Ferrier2016,QD-PRL,binary-QD}, quantum droplets have attracted 
considerable attention as dilute quantum liquids exhibiting properties beyond those of conventional Bose-Einstein condensates (BEC) \cite{Cavicchioli2025,Ferioli2020,Ferioli2019,Dong2021,Ma2023}. In this context, a recent 
study introduced the concept of elastic modulus into the investigation of three-dimensional quantum droplets, where the corresponding quantity can be identified 
as the bulk modulus \cite{3D-Bulkmodulus}. Based on a uniform-density approximation, the bulk modulus was evaluated from the response of the chemical potential to a change in the 
droplet volume and was further related to the breathing-mode frequency. This connection links the collective dynamics of quantum droplets to their intrinsic mechanical 
response, providing a useful perspective for characterizing self-bound quantum fluids.

However, QDs may exhibit qualitatively different properties in different spatial dimensions, because the form and role of the beyond-mean-field correction 
depend strongly on dimensionality \cite{LHY-1D,LZH-FOP}. In three dimensions, the LHY correction is repulsive and counteracts the collapse driven by residual mean-field attraction, thereby 
stabilizing liquid-like droplets \cite{kartashov-3D,Dong-3D,Ma-shell,Otajonov-3D,Guo-LHY}. In one dimension, by contrast, the stabilization mechanism 
is essentially reversed: the LHY-induced quadratic nonlinear term acts effectively 
as an attractive self-focusing contribution, while the mean-field interaction is repulsive \cite{Astrakharchik-dynamical,Parisi-QDs-1D,Mistakidis-formation,Tylutki-collective}. 
As a result, one-dimensional QDs are not simply reduced versions of 
their three-dimensional counterparts, but are supported by a different balance of nonlinearities \cite{Pylak-crossover,Ilg-crossover}. Two-dimensional droplets also show their 
own dimensional peculiarity, since the LHY correction has a logarithmic density dependence \cite{Lin-2d,Li-2dvortex,Dong-Internal,Hu-Collisional,Stumer-2D,Examilioti-2D}.

A particularly important feature of one-dimensional systems is that they can support stable bright solitons \cite{Pitaevskii-BEC-book,Boris-Multidimensional-soliton}. As the particle number increases, such systems may gradually 
evolve from a strongly localized soliton-like state to a droplet-like state with a broader density profile and an approximately saturated central density. This smooth 
soliton-to-droplet crossover is reflected in several equilibrium properties, such as the chemical potential, peak density, and characteristic width, which provide direct 
information on how the self-bound state approaches the liquid-like regime \cite{Mithun-Statistical,Otajonov-stationary,Edmonds-QDs,ZhaoF-Discrete,Morera-universal,Charalampidis-2component}. Beyond these static quantities, 
the breathing oscillation offers a dynamical probe of the compressional 
response of the system \cite{Katsimiga-interaction,Lv-Breather,Depalo-quasi-1D,DuX-ground}. Therefore, one-dimensional QDs provide a natural setting for extending 
the concept of elastic modulus to low-dimensional self-bound quantum fluids 
and for examining how the modulus is connected to the breathing-mode frequency in comparison with the three-dimensional case.

In this work, we study the elastic modulus of one-dimensional quantum droplets. Using a super-Gaussian variational approach for analytical calculations and the spatial 
scaling method for numerical simulations, we determine the elastic modulus from both analytical and numerical perspectives. We further examine its relation to the eigenfrequency 
of the breathing mode, thereby extending the modulus-frequency connection previously established for three-dimensional droplets to the one-dimensional case. In contrast to the three-dimensional setting, 
the existence of stable soliton-like states in one dimension makes the dependence of the ratio $\eta=B/\Omega^2$ on the interaction strength $g$ and particle number $N$ less reducible to a simple power-law scaling.
The subsequent 
presentation is organized as follows. In Sec.~\ref{model}, we introduce the theoretical model. In Sec.~\ref{The Variational Approximation For The ground states}, we perform a variational analysis of the ground-state solution. 
In Sec.~\ref{Comparison of Numerical and VA Results}, we compare the numerical results with the variational predictions. In Sec.\ref{The relation between bulk modulus and frequency}, we discuss the relation between the elastic modulus and the breathing-mode frequency. 
Finally, Sec.~\ref{Conclusion} summarizes the main findings of this work.
 
\section{The model}  \label{model}

The dynamics of Bose-Einstein condensate (BEC) in one-dimensional (1D) free space is described by the dimensionless Gross-Pitaevskii equation (GPE) for the wave function 
$\psi$. This equation incorporates both the cubic mean-field interaction and the Lee-Huang-Yang (LHY) correction, which take the quadratic form in 1D \cite{LHY-3d,LHY-1D}:
\begin{equation}
	i\frac{\partial}{\partial t}\psi=-\frac{1}{2}\frac{\partial^{2}}{\partial x^2}\psi+g|\psi|^2\psi-|\psi|\psi,\label{dimensionless-GPE}
\end{equation}%
where $g>0$ is the repulsive attraction strength.
The total atom number in the system is
\begin{equation}
	N=\int |\psi|^2dx.
\end{equation}%
The corresponding Hamiltonian (total energy) of the system corresponding to Eq. (\ref{dimensionless-GPE}) is
\begin{equation}
	E=\int\left( \frac{1}{2}|\partial_x \psi|^2+\frac{1}{2}g|\psi|^4-\frac{2}{3}|\psi|^3\right)dx.\label{energy}
\end{equation}%
To find stationary solutions, we adopt the standard ansatz 
$\psi (\mathbf{x},t)=\phi (\mathbf{x})e^{-i\mu t}$, where $\mu$ denotes the chemical potential and $\phi $
is the time-independent wave function. Substituting this ansatz into Eq. (\ref{dimensionless-GPE}) yields the stationary GPE:
\begin{equation}
	\mu\phi=-\frac{1}{2}\frac{\partial^{2}}{\partial x^2}\phi+g|\phi|^2\phi-|\phi|\phi.
\end{equation}%

In this work, we investigate the elastic modulus $B$ of these quantum droplets, which is defined as \cite{BEC-bulk,3D-Bulkmodulus}:
\begin{equation}
	B=-W\frac{\partial p}{\partial W}=W\frac{\partial^2 E}{\partial W^2},\label{Bulkmodulus}
\end{equation}%
where $p=-\frac{\partial E}{\partial W}$ is the effective one-dimensional pressure, $E$ is the total energy from Eq. (\ref{energy}), and $W=(\int |\phi|^2dx)^2/\int |\phi|^4dx$ is the effective length of the droplet.
The two independent control parameters in our study are the total atom number $N$ and the contact interaction strength \(g\).

\section{The Variational Approximation For The ground states}
\label{The Variational Approximation For The ground states}
\begin{figure}[tbp]
	\centering
	{\includegraphics[width=0.85\columnwidth]{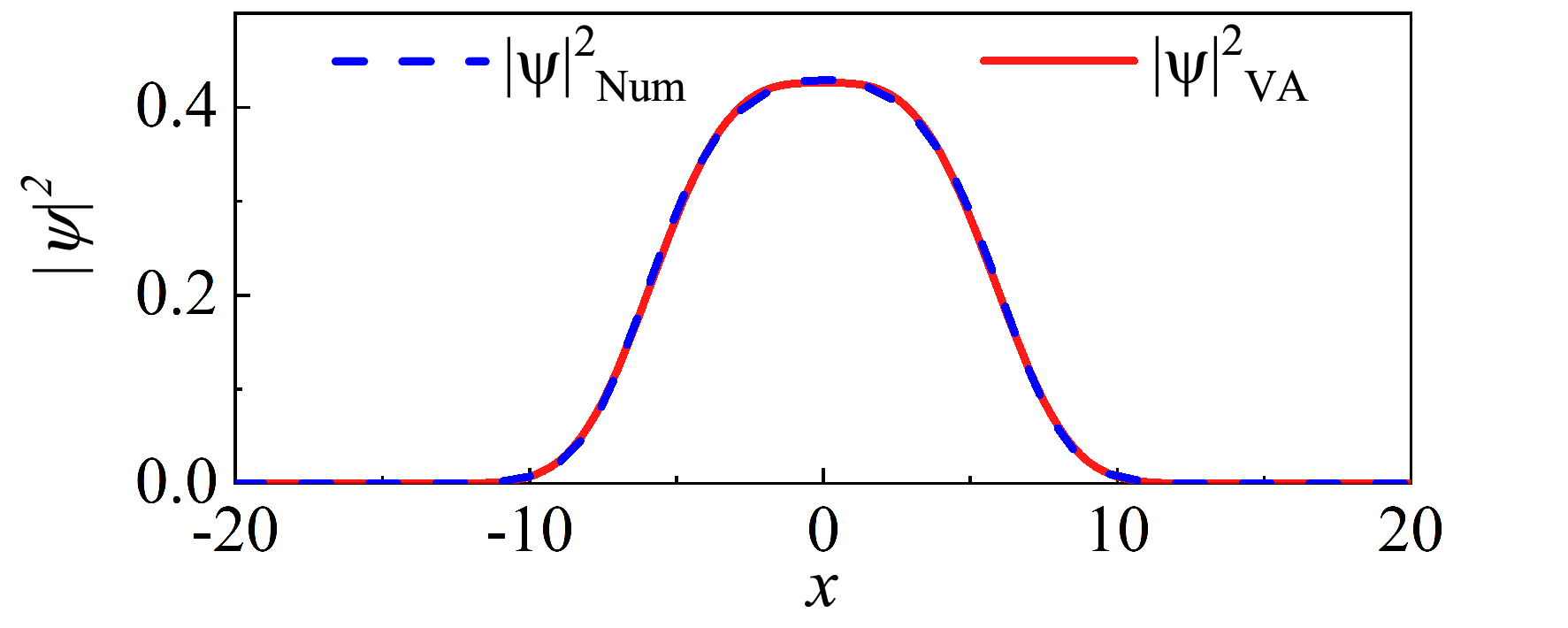}}
	\caption{Radial density distribution of stationary wave function \(\phi\) with $N = 5$ and $g=1$ . 
		The red solid line represents the numerical result $|\phi_{\mathrm{VA}}|^2$, and blue dashed line denotes the variational result $|\phi_{\mathrm{Num}}|^2$,
		obtained using the VA parameter \(w=6.54\) and \(\alpha=1.66\).}
	\label{VA-Num-pk}
\end{figure}

The Lagrangian density corresponding to Eq. (\ref{dimensionless-GPE}) is
\begin{equation}
	\mathcal{L}=\frac{i}{2}(\psi^*\partial_t\psi-\psi\partial_t\psi^*)-\frac{i}{2}|\partial_x \psi|^2-\frac{1}{2}g|\psi|^4+\frac{2}{3}|\psi|^3,
\end{equation}%
and the corresponding Lagrangian is 
\begin{equation}
	L=\int \mathcal{L}dx.
	\label{Lagrangian-1}
\end{equation}
The variational approximation (VA) \textit{ans\"{a}tze} for 1D quantum droplets
are denoted as super-Gaussian \cite{Otajonov-stationary,ft-bec}
\begin{equation}
	\psi(x,t)=A\exp\left[-\frac{1}{2}\left(\frac{x}{w}\right)^{2\alpha}+i\beta x^2\right],
	\label{ansatz}
\end{equation}%
where \(A(t), w(t)\), and \(\beta(t)\) are variational parameters representing the amplitude, width, and chirp of the super-Gaussian, respectively.
\(\alpha\) denotes the order of the super-Gaussian, and we assume \(\alpha\) remains constant over time. 
The amplitude \(A\) can be determined through normalization conditions,
\begin{equation}
	N=\int |\psi|^2 dx=\frac{w}{\alpha}\Gamma(\frac{1}{2\alpha})A^2,
\end{equation}%
The effective length of the QDs is:
\begin{equation}
	W=\frac{(\int|\psi|^2dx)^2}{\int|\psi|^4dx}=\frac{2^{\frac{1}{2\alpha}}\Gamma(\frac{1}{2\alpha})}{\alpha}w=C_0w,\label{W1}
\end{equation}%
Here $C_0=\frac{2^{\frac{1}{2\alpha}}\Gamma(\frac{1}{2\alpha})}{\alpha}$. The substitution of the super-Gaussian ansatz (\ref{ansatz}) 
in the Lagrangian (\ref{Lagrangian-1}) yields the Lagrangian
\begin{equation}
	L=\int\mathcal{L} dx
	=-C_1N\dot{\beta}w^2-\frac{C_1C_2}{2}\frac{\alpha N}{w^2} -2C_1N\left(\beta w\right)^2-gC_1C_3\frac{N^2}{w}-2C_1C_4\frac{N^{\frac{3}{2}}}{w^{\frac{1}{2}}},
	\label{Lagrangian-2}
\end{equation}
with coefficients $C_i,(i=1,2,3,4)$ given by
\begin{equation}
	C_1=\frac{\Gamma(\frac{3}{2\alpha})}{\Gamma(\frac{1}{2\alpha})},
	C_2=\frac{\alpha\Gamma(2-\frac{1}{2\alpha})}{\Gamma(\frac{3}{2\alpha})},
	C_3=\frac{\alpha}{2^{1+\frac{1}{2\alpha}}\Gamma(\frac{3}{2\alpha})},
	C_4=-\frac{2^{\frac{1}{2\alpha}}\alpha^\frac{1}{2}\Gamma^{\frac{1}{2}}(\frac{1}{2\alpha})}{3^{1+\frac{1}{2\alpha}}\Gamma(\frac{3}{2\alpha})}.
	\label{C-matrix}
\end{equation}
For chirp $\beta$, the Euler-Lagrangian equations 
leads to the following equation:
\begin{equation}
	\beta=\frac{\dot{w}}{2w},
	\label{Eoula-Eqs-beta}
\end{equation}%
Substituting (\ref{Eoula-Eqs-beta}) into Lagrangian (\ref{Lagrangian-2}) yields an effective Lagrangian without $\beta$
\begin{equation}
	L=-\frac{C_1}{2}Nw\ddot{w}-\frac{C_1C_2}{2}\frac{\alpha N}{w^2}
	-gC_1C_3\frac{N^2}{w}
	-2C_1C_4\frac{N^{\frac{3}{2}}}{w^{\frac{1}{2}}}.\label{EFF-Lagrangian}
\end{equation}
The Euler-Lagrangian equation for width $w$ is
\begin{equation}
	\ddot{w}=C_2\frac{\alpha}{w^3}+gC_3\frac{ N}{w^2}+ C_4\frac{N^{\frac{1}{2}}}{w^{\frac{3}{2}}}\equiv -\frac{dU}{dw},\label{Eoula-Eqs-W}
\end{equation} 
where \(U\) corresponds to an effective potential. 

To obtain the intrinsic oscillation frequency $\Omega$, we consider small-amplitude oscillations of the width around its equilibrium 
value, expressed as $w=w_0+\delta w$, where $w_0$ denotes the equilibrium width and $\delta w \ll w_0$. Substituting this ansatz into 
Eq.~(\ref{Eoula-Eqs-W}) and performing linearization leads to
\begin{equation}
	\delta\ddot{ w}+\Omega^2\delta w=0.
	\label{oscillation-eq} 
\end{equation}
Here, 
\begin{equation}
	\Omega^2_{\text{VA}}=\frac{d^2U}{dw^2}=3C_2\frac{\alpha}{w^4}+2gC_3\frac{ N}{w^3}
	+\frac{3C_4}{2}\frac{N^{\frac{1}{2}}}{w^{\frac{5}{2}}},
	\label{Omega2}
\end{equation}
represents the squared eigenfrequency of the intrinsic oscillation.

\begin{figure}[tbp]
	\centering
	{\includegraphics[width=0.8\columnwidth]{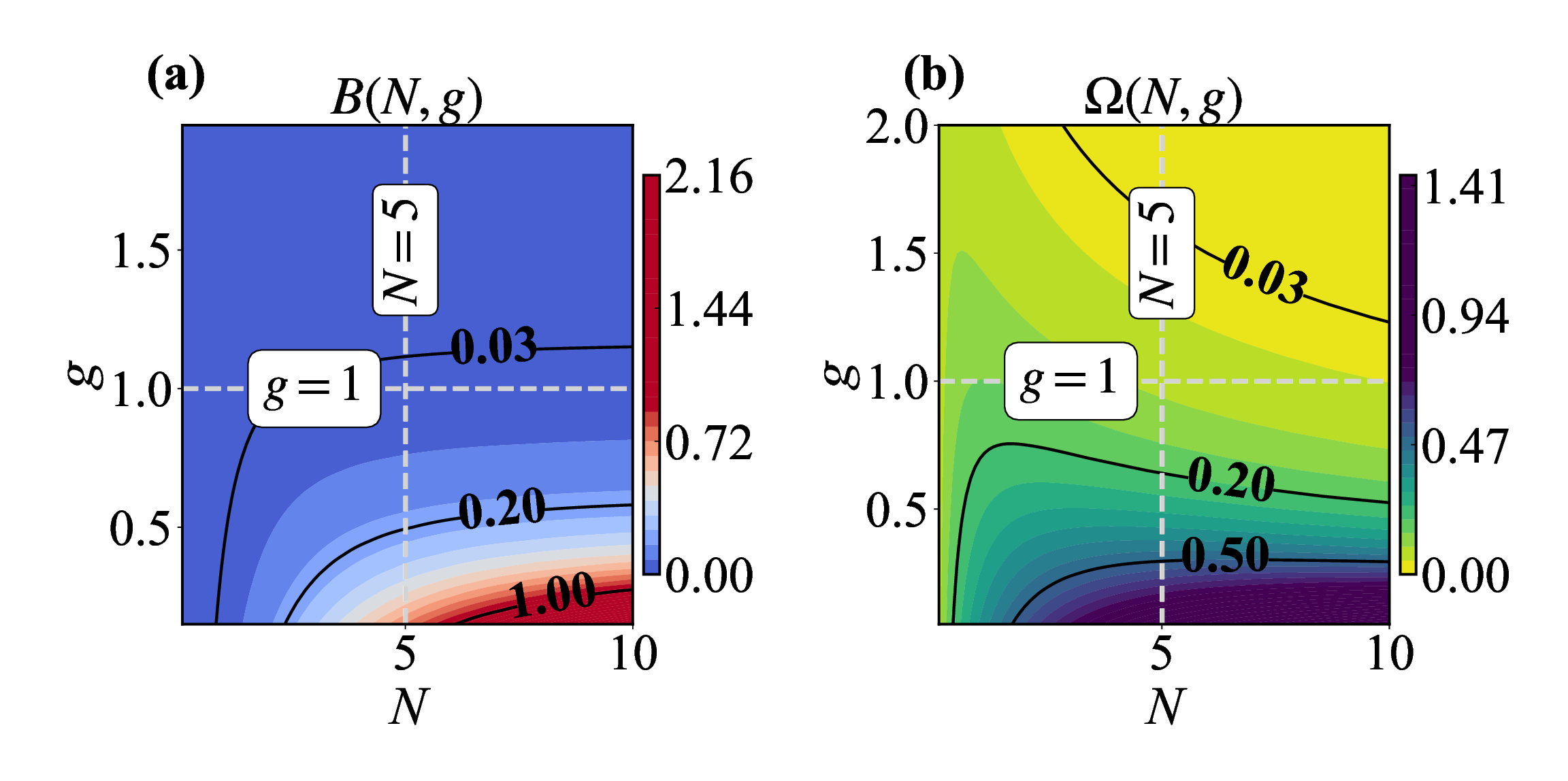}} %
	\caption{Density plots of (a) the elastic modulus $B_{\mathrm{VA}}$ and (b) the breathing-mode frequency 
	$\Omega_{\mathrm{VA}}$ in the $(N,g)$ plane. The color scale from red to blue in panel (a) denotes increasing 
	values of $B_{\mathrm{VA}}$, whereas darker colors in panel (b) correspond to larger values of 
	$\Omega_{\mathrm{VA}}$. The black curves indicate the isovalue contours in panels (a) and (b). 
	The white dashed lines mark $g=1$ and $N=5$, corresponding to the curves shown in Figs.~\ref{Bomega-gN}(a,b) and (c,d), respectively. }
	\label{B_Omega_N_g}
\end{figure}

For stationary solutions, the equilibrium width is determined by setting $\ddot{w}=0$ in Eq.~(\ref{Eoula-Eqs-W}), whereas the super-Gaussian order $\alpha$ 
follows from the variational condition $\partial L/\partial \alpha=0$. Solving the resulting coupled equations yields the stationary variational 
parameters $w$ and $\alpha$. Fig.~\ref{VA-Num-pk} presents a comparison between the density profiles obtained from the VA and 
numerical simulations for a quantum droplet with $N=5$ and $g=1$. The corresponding variational parameters are found to be $w=6.54$ and $\alpha=1.66$. The value $\alpha=1.66>1$ indicates the necessity of employing a super-Gaussian ansatz to capture the flat-top density profile of the droplet. And the excellent agreement between the numerical and variational results further validates the accuracy of the VA.

Furthermore, according to Eq. (\ref{energy}), the energy \(E\) is obtained as
\begin{equation}
	E_\text{VA}=N\left[\frac{C_1C_2}{2}\frac{\alpha}{w^2}+gC_1C_3\frac{N}{w}
	+2C_1C_4\left(\frac{N}{w}\right)^{\frac{1}{2}}\right],\label{nengliangVA}
\end{equation}
and the chemical potential is
\begin{equation}
	\mu_\text{VA}=\frac{d E_\text{eq}}{d N}=\frac{C_1C_2}{2}\frac{\alpha}{w^2}+2gC_1C_3\frac{ N}{w} +3C_1C_4\frac{N}{w}^{\frac{1}{2}},
	\label{chemical potential}
\end{equation}
where $E_\text{eq}$ is the equilibrium energy, determined by the equilibrium condition $\left( \partial E / \partial w \right)_N = 0$.
Accordingly, the modulus \(B\) can obtained as
\begin{equation}
	\begin{split}
		B_\text{VA}=\frac{N}{W}\left(3C_1C_2\frac{\alpha}{w^2}+2gC_1C_3\frac{N}{w} +\frac{3C_1C_4}{2}\frac{N}{w}^{\frac{1}{2}}\right),\label{VA-Bulkmodulus}
	\end{split}
\end{equation}

According to Eqs.~(\ref{Omega2}) and (\ref{VA-Bulkmodulus}), we plot the  
frequency \(\Omega(N, g)\) and the modulus \(B(N, g)\) in 
Figs.~\ref{B_Omega_N_g}(a,b). It can be seen that as the particle number \(N\) increases, 
the frequency \(\Omega\) first increase and then decreases, with a maximum value, while the modulus \(B\) 
increases. In contrast, as \(g\) increases---corresponding to a stronger 
repulsive interaction---both \(\Omega\) and \(B\) decreases.

\begin{figure}[t]
	\centering
	{\includegraphics[width=0.65\columnwidth]{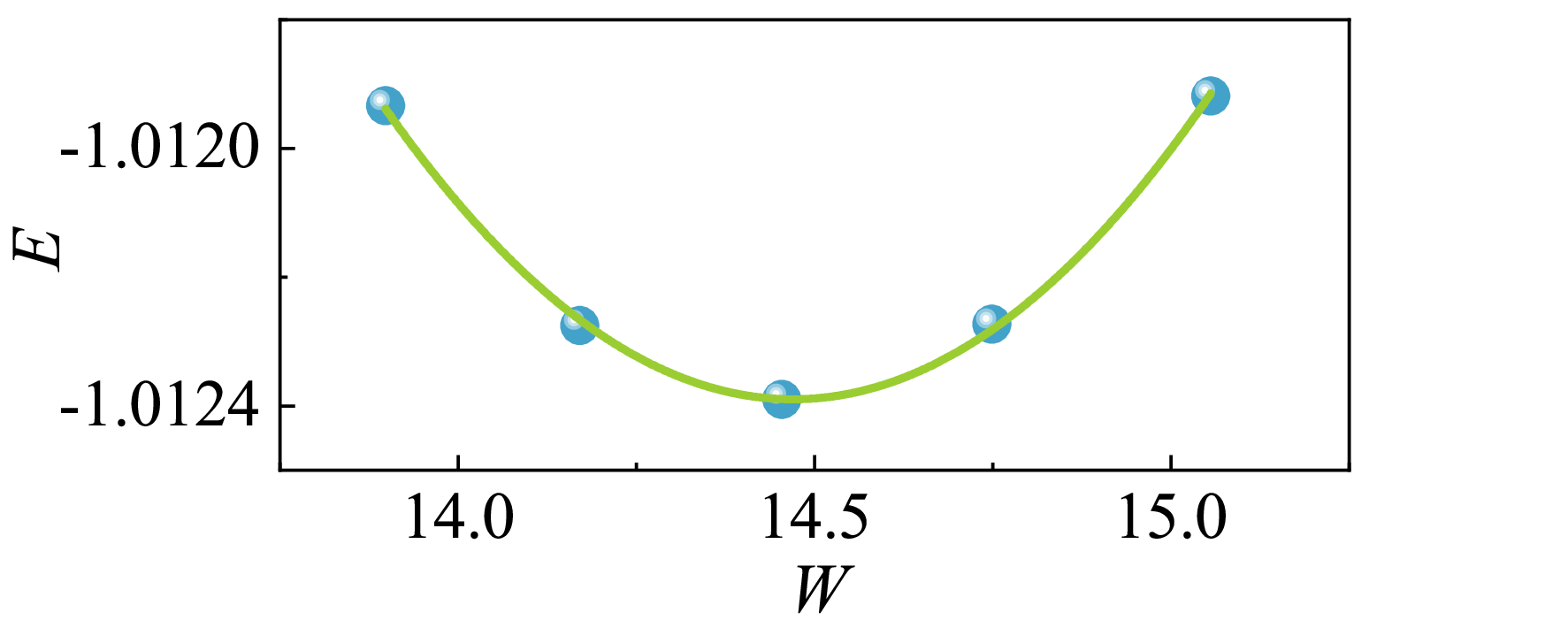}}
	\caption{Energy-effective-length relation of a quantum droplet with $N=5$ and $g=1$ under different spatial-scaling factors $a=0.96, 0.98, 1.00, 1.02,$ and $1.04$. The blue filled circles denote the numerical results obtained using the spatial-scaling method. The green solid line represents the quadratic polynomial fit, $E=0.0014W^2-0.04W-0.72$.}
	\label{N-30w-evolution}
\end{figure}

\section{Comparison of Numerical and VA Results}
\label{Comparison of Numerical and VA Results}
In the numerical simulations, to obtain the frequency $\Omega$ and the bulk modulus $B$ of quantum droplets (or solitons), we perform simulations using the 
imaginary-time propagation (ITP) method and the BdG approach \cite{ITP}. The detailed procedure is as follows:

\begin{enumerate}
	\item \textbf{Ground-state preparation:} 
	For given parameters \((N, g)\), we solve Eq.~(\ref{dimensionless-GPE}) using the ITP to obtain a stable ground-state QD.
	
	\item \textbf{Spatial scaling method for extracting the modulus \(B\):}
	To numerically extract the 
	elastic modulus of the quantum droplet, we employ a spatial scaling transformation defined by \(x' = a x\). To preserve normalization, the wave function 
	is rescaled as \(\psi' = \psi / \sqrt{a}\). This transformation effectively introduces a controlled deformation of the system, where the parameter 
	\(a\) characterizes the strength of the transformation and satisfies \(a \approx 1\).
	Under this scaling, the quantum droplet acquires a corresponding effective width \(W\) and energy \(E\) for each value of \(a\). By varying \(a\) in 
	the vicinity of unity, we construct the relation \(E(W)\) (Eq. (\ref{energy}) and (\ref{W1})), which allows us to extract the curvature 
	\(\partial^2 E / \partial W^2\), and thereby determine the modulus \(B\).
	Furthermore, as demonstrated in Appendix C, the modulus \(B\) is invariant under the spatial scaling transformation when $\alpha\approx1$, ensuring the consistency and 
	physical validity of the method.
	
	\item \textbf{Bogoliubov-de Gennes (BdG) approach:}
	We extract the oscillation frequency using the Bogoliubov-de Gennes method. A small perturbation is introduced to the stationary state, 
	and the resulting ansatz is substituted into the Eq. (\ref{dimensionless-GPE}). Linearizing with respect to the perturbation amplitude leads 
	to a linear eigenvalue problem. The eigenfrequencies \( \Omega \) are obtained from the resulting BdG equations, and the lowest non-zero 
	real eigenvalue is identified as the fundamental oscillation frequency of the system.
\end{enumerate}

\begin{figure}[htbp]
	\centering
	{\includegraphics[width=0.9\columnwidth]{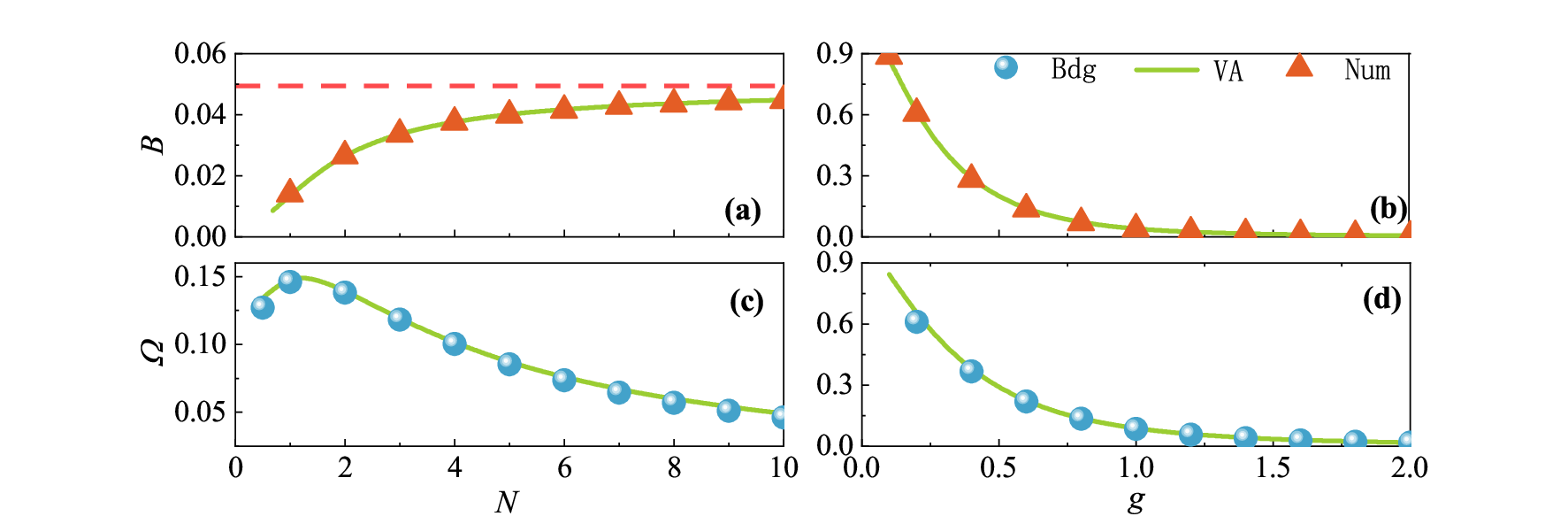}}
	\caption{(a--b) The modulus $B$ versus 
		$g$ and $N$. The green solid lines represent the results obtained from the VA [Eq.~(\ref{VA-Bulkmodulus})]. 
		The red triangles denote the numerical results. The red dashed line denotes the asymptotic value $B_\infty=4/(81g^3)\simeq 0.0494$.
		(c--d) The frequency $\Omega$ 
		as a function of $g$ and $N$. The green solid lines show the VA 
		[Eq.~(\ref{Omega2})], the blue spherical dots denote the BdG results.In panels (a) and (c), 
		we fix \(g=1\), while in panels (b) and (d), we fix $N=5$.}
	\label{Bomega-gN}
\end{figure}

This procedure enables a direct comparison between the VA and the numerical results. For a quantum droplet 
with \(N=5\) and \(g=1\), the modulus \(B\) are determined using the methods described above. Fig.~\ref{N-30w-evolution} illustrates the 
dependence of the energy \(E\) on the effective length \(W\), obtained using the spatial scaling method. The blue circles denote the 
numerical data corresponding to the scaling factors \(a=0.96,\,0.98,\,1.00,\,1.02,\) and \(1.04\), while the green solid curve represents 
the quadratic fit to these data. From the fitted \(E(W)\) relation, the modulus is evaluated through Eq.~(\ref{Bulkmodulus}), yielding 
\(B_{\mathrm{Num}}=0.04\). This value is in excellent agreement with the variational prediction \(B_{\mathrm{VA}}=0.04\), thereby validating 
both the variational approximation and the spatial scaling method employed for extracting the modulus.

Therefore, employing this methodology, we compare the numerical results of the modulus and oscillation frequency with the VA predictions in Fig.~\ref{Bomega-gN}. 
Figs.~\ref{Bomega-gN}(a,c) correspond to fixed $N=5$, whereas Figs.~\ref{Bomega-gN}(b,d) correspond to fixed $g=1$; the modulus is shown in Figs.~\ref{Bomega-gN}(a,b), and the oscillation frequency in Figs.~\ref{Bomega-gN}(c,d). 
For the modulus, the green solid lines represent the VA results given by Eq.~(\ref{VA-Bulkmodulus}), while the red triangles denote the numerical results obtained by the above procedure. 
The modulus $B$ increases with the particle number $N$ and gradually approaches the asymptotic value $B_{\infty}=4/(81g^3)$, as given by Eq.~(\ref{Blim}); in particular, for $g=1$, one has $B_{\infty}=4/81\simeq 0.0494$. 
In contrast, increasing the repulsive interaction strength $g$ suppresses the modulus, driving it toward zero. 
For the oscillation frequency, the VA predictions given by Eq.~(\ref{Omega2}) are compared with the BdG results obtained from the lowest nonzero real eigenfrequency, showing good agreement in Figs.~\ref{Bomega-gN}(c,d). 
The frequency first increases and then decreases, eventually tending to zero as the particle number $N$ increases, while increasing the repulsive interaction strength $g$ also suppresses $\Omega$ toward zero.

\begin{figure}[htbp]
	\centering
	{\includegraphics[width=0.75\columnwidth]{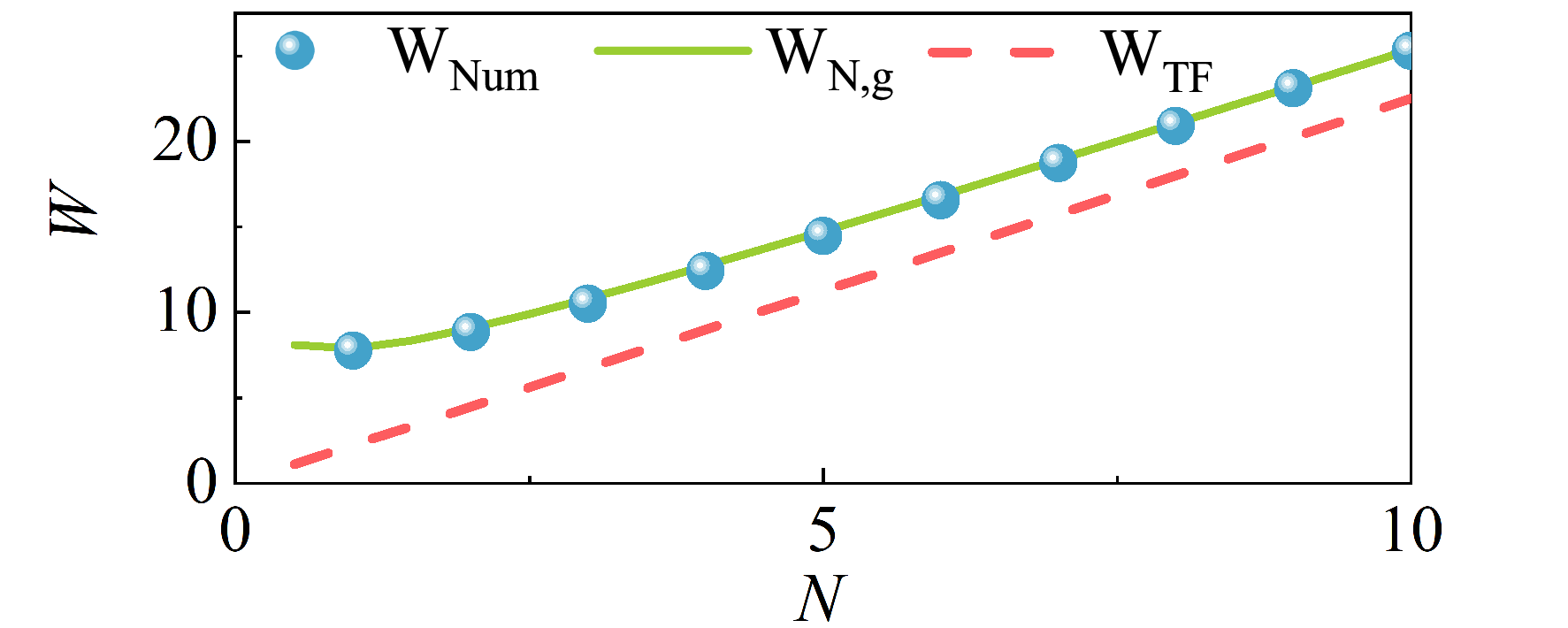}}
	\caption{Effective length $W$ as a function of the particle number $N$ for QDs with $g=1$. The green solid line denotes the VA prediction $W_{N,g}$, the 
	blue circles represent the numerical results, and the red dashed line denotes the Thomas-Fermi (TF) approximation result $W_{\mathrm{TF}}$.}
	\label{W}
\end{figure}

\section{The relation between bulk modulus \(B\) and frequency \(\Omega\)}
\label{The relation between bulk modulus and frequency}
\begin{figure}[htbp]
	\centering
	{\includegraphics[width=0.9\columnwidth]{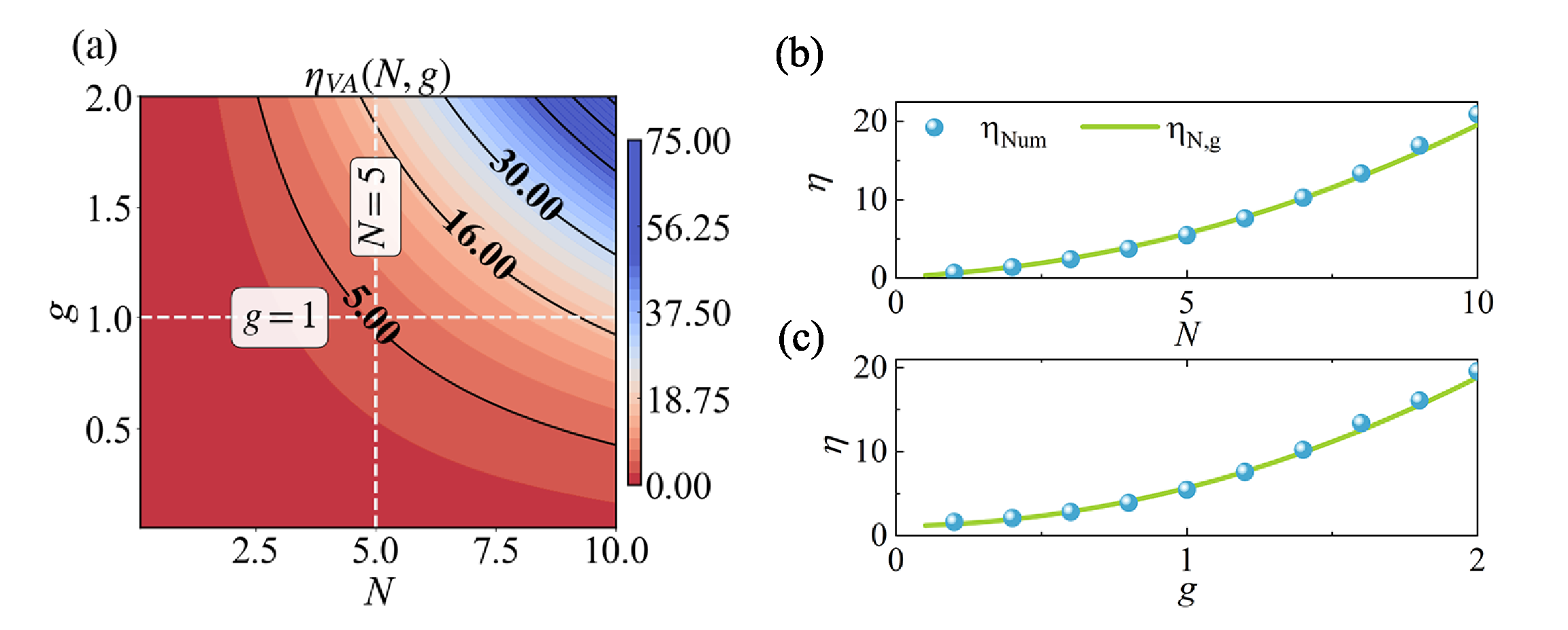}}
	\caption{Panel (a) presents 
		the $\eta_{\mathrm{VA}}(N,g)$ density map, where the horizontal axis denotes the particle number $N$ and interaction strength 
		$g$. The black contour lines indicate the isovalues of $\eta_{\mathrm{VA}}$, with $\eta_{\mathrm{VA}}$ increasing from bottom to 
		top. The two white dashed lines correspond to $g=1$ and $N=5$, which are associated with 
		panels (b) and (c), respectively. Panels (b) and (c) show the dependence of $\eta$ on $g$ and $N$, respectively, with (b) corresponding to 
		the fixed interaction strength 
		$g=1$ and (c) to the fixed particle number $N=5$. In both panels, the green solid curves denote $\eta_{N,g}$, given by Eq.~(\ref{eta_eqs}),and the blue spherical dots denote the numerical results obtained from $\eta_{\mathrm{Num}}=B_{\mathrm{Num}}/\Omega_{\mathrm{Bdg}}^2$ .}
	\label{eta_N_g}
\end{figure}

In classical mechanics, the elastic modulus is usually related to the square of the oscillation frequency~\cite{Rao-mechanical},
\begin{equation}%
	B \varpropto \Omega^2. \label{class-B-Omega2}
\end{equation}%
Motivated by this analogy, we examine whether a similar relation exists between the bulk modulus \(B\) and the breathing frequency \(\Omega\) of quantum droplets. To this end, we define the ratio
\begin{equation}
	\begin{split}
		\eta=\frac{B}{\Omega^2}. \label{ETA}
	\end{split}
\end{equation}%
Substituting Eqs.~(\ref{Omega2}) and (\ref{VA-Bulkmodulus}) into Eq.~(\ref{ETA}), we obtain
\begin{equation}
	\eta_\text{VA}=\kappa N W_\text{VA}, \label{eta_relation}
\end{equation}%
where $\kappa=C_1/C_0^2$. As shown in Fig.~\ref{eta_N_g}(a), \(\eta_\text{VA}\) increases with both the particle number \(N\) and the interaction strength 
\(g\), indicating that, within the VA framework, the proportionality coefficient between \(B\) and \(\Omega^2\) is not a universal constant, but is 
determined by the system parameters \(N\) and \(g\).

To make this dependence explicit, we further express the VA equilibrium width in terms of \(N\) and \(g\). As derived in Appendix~E, it can be approximated by
\begin{equation}
	\begin{split}
		W_{N,g}=\frac{9}{4}Ng^2+\frac{27}{4}\left(\frac{C_2\alpha C_4^2}{4C_3^3}\right)^{2/3}N^{-1/3}+\frac{1}{2}g. \label{WTF}
	\end{split}
\end{equation}%
As shown in Fig.~\ref{W}, we compare \(W_{\rm Num}\), \(W_{N,g}\), and \(W_{\rm TF}\) (\ref{TF-approximate}, Eq.~(\ref{W-Tf})). 
The expression \(W_{N,g}\) [Eq.~(\ref{WTF})] agrees well with the numerical results over the considered parameter range, while 
the Thomas-Fermi (TF) prediction \(W_{\mathrm{TF}}\) exhibits a noticeable deviation, particularly in the small-\(N\) region 
where the TF approximation becomes less accurate. Meanwhile, \(W_{N,g}\) asymptotically approaches \(W_{\mathrm{TF}}\) as \(N\to\infty\), 
showing that the TF result is recovered in the large-particle-number limit. This comparison further confirms the validity of \(W_{N,g}\). Therefore, replacing 
\(W_\text{VA}\) in Eq.~(\ref{eta_relation}) by Eq.~(\ref{WTF}), we arrive at
\begin{equation}
	\begin{split}
		\eta_{N,g}
		=\kappa N\left[
		\frac{9}{4}Ng^2+\frac{27}{4}\left(\frac{C_2\alpha C_4^2}{4C_3^3}\right)^{2/3}N^{-1/3}+\frac{1}{2}g
		\right]. \label{eta_eqs}
	\end{split}
\end{equation}%

Eq.~(\ref{eta_eqs}) indicates that the ratio \(\eta=B/\Omega^2\) is jointly governed by the particle number \(N\) and the interaction strength \(g\). 
Thus, although \(B\) and \(\Omega^2\) are closely correlated, their relation is not characterized by a constant proportionality coefficient over the whole parameter region. 
As shown in Figs.~\ref{eta_N_g}(b,c), we compare \(\eta_{\rm Num}=B_{\rm Num}/\Omega_{\rm BdG}^2\) with \(\eta_{N,g}\), and find that they agree well within the selected parameter region, confirming the accuracy of Eq.~(\ref{eta_eqs}). 
This behavior is different from the three-dimensional case, where the corresponding ratio follows a simpler power-law dependence on the control parameters. In one dimension, 
the presence of stable soliton-like states introduces finite-particle-number corrections, leading to a more involved dependence of \(\eta\) on \(g\) and \(N\).
This relation also suggests that, in experiments, the bulk modulus of quantum droplets can be inferred by measuring the corresponding oscillation frequency.

\section{Conclusion}
\label{Conclusion}
In this work, we perform a variational analysis of the dimensionless one-dimensional Gross-Pitaevskii equation (GPE) for quantum droplets, 
based on a super-Gaussian trial function. Analytical expressions for the droplet's oscillation frequency \(\Omega\), chemical potential \(\mu\), 
and effective length \(W\) are derived, from which the modulus is obtained. 
By comparing the variational approximation (VA) results with direct numerical simulations, we confirm the validity of the super-Gaussian-based VA.

In the numerical simulations, we use the spatial scaling method to obtain the energy curve \(E(W)\) around the equilibrium state. 
The curvature \(\partial^2 E/\partial W^2\) is then extracted from a quadratic fit, and the bulk modulus is calculated as
\[B=-W\frac{\partial p}{\partial W}
=W\frac{\partial^2 E}{\partial W^2}.\]
The oscillation frequency \(\Omega\) is obtained independently from the BdG analysis. 
Both the numerical modulus and the BdG frequency are in good agreement with the VA predictions. For fixed \(g\), the modulus \(B\) 
increases with the particle number \(N\) and gradually approaches the asymptotic value \(B_{\infty}=4/(81g^3)\); in particular, for \(g=1\), 
one has \(B_{\infty}=4/(81g^3)\approx 0.0494\). Meanwhile, the oscillation frequency \(\Omega\) first increases and then decreases with increasing \(N\), 
eventually tending to zero in the large-\(N\) limit. Conversely, for fixed \(N\), increasing the repulsive interaction strength \(g\) suppresses 
both \(B\) and \(\Omega\), driving them gradually toward zero.

Finally, inspired by the analogy with classical mechanics, where the elastic modulus is correlated with the vibration frequency, we define the characteristic parameter
\(\eta=B/\Omega^{2}\). Within the VA framework, we obtain \(\eta_{\mathrm{VA}}\) and find that it increases with both the particle number \(N\) and the interaction strength \(g\). 
To further establish the explicit dependence of \(\eta\) on the control parameters \(N\) and \(g\), we combine the variational expressions of \(B\) and \(\Omega^2\), leading 
to an analytical relation \(\eta_{N,g}\). The comparison with numerical results confirms the accuracy of this analytical expression. Compared with the three-dimensional case, 
the stable existence of solitons in one dimension makes \(\eta\) exhibit a more complex dependence on \(N\) and \(g\).

This relation also provides practical guidance for experimentally determining the bulk modulus \(B\). 
For sufficiently large particle numbers, \(B\) can be directly estimated from its asymptotic expression once the interaction strength \(g\) is known. 
Alternatively, \(B\) can be inferred from the measured oscillation frequency through the relation \(\eta=B/\Omega^{2}\).

The present work may be extended in several directions. One possible extension is to investigate the elastic response of low-dimensional quantum 
droplets with dipole-dipole interactions. Since the dipole-dipole interaction is intrinsically anisotropic, such systems may exhibit direction-dependent 
mechanical properties, making it possible to define and analyze anisotropic elastic moduli of quantum droplets \cite{Sinha-quasi-1D,LiG-anisotropic,Li-2D}. This 
would also allow one to explore the 
relation between the elastic response and internal collective excitations, such as breathing, quadrupole, and other deformation modes 
\cite{CaiY,Ground-state,Bisset-Ground,Petter-Probing,DinizPC,Baillie-collective,ZjangF,Ferrier-Scissors,Blakie-axial}. 
Another interesting 
direction is to study quantum droplets in optical lattices \cite{Jaksch-OL,Edwards-VA-ol}. The presence of a periodic potential may modify the equilibrium density distribution, effective 
compressibility, and collective excitation spectrum, thereby providing a possible route to controlling the elastic modulus of quantum droplets 
\cite{Nie-OL,QDs-1d-OL,Morera-universal,ZhouZ-dynamical-ol,ZhouZ-Controllable,Ancilotto-vortex-ol}. 
These extensions would further deepen the understanding of the mechanical properties of self-bound quantum fluids in more complex settings.

\appendix

\section{The Thomas-Fermi (TF) approximation}
\label{TF-approximate}

The energy $E_{\rm TF}$ within the TF approximation is given by
\begin{equation}
	\begin{split}
		E_{\rm TF}
		=\int \left(\frac{1}{2}gn^2-\frac{2}{3}n^{\frac{3}{2}}\right)dx
		=\frac{1}{2}gnN-\frac{2}{3}n^{\frac{1}{2}}N .
	\end{split}
\end{equation}
The equilibrium density is determined by minimizing $E_{\rm TF}$ with respect to $n$,
\begin{equation}
	\begin{split}
		\frac{dE_{\rm TF}}{dn}
		=\frac{1}{2}gN-\frac{1}{3}n^{-\frac{1}{2}}N=0 .
	\end{split}
\end{equation}
This yields the equilibrium particle density
\begin{equation}
	\begin{split}
		n_e=\frac{4}{9g^2}.
	\end{split}
\end{equation}
Therefore, the effective length in the TF approximation is
\begin{equation}
	\begin{split}
		W_e=\frac{N}{n_e}=\frac{9}{4}Ng^2 .
		\label{W-Tf}
	\end{split}
\end{equation}

To obtain the bulk modulus in the TF limit, we rewrite the TF energy as a function of the effective length $W$ by using $n=N/W$:
\begin{equation}
	\begin{split}
		E_{\rm TF}(W)
		=\frac{1}{2}g\frac{N^2}{W}
		-\frac{2}{3}\frac{N^{3/2}}{W^{1/2}} .
	\end{split}
\end{equation}
The corresponding effective pressure is
\begin{equation}
	\begin{split}
		p=-\frac{\partial E_{\rm TF}}{\partial W}
		=\frac{1}{2}g\frac{N^2}{W^2}
		-\frac{1}{3}\frac{N^{3/2}}{W^{3/2}} .
	\end{split}
\end{equation}
Accordingly, the bulk modulus is defined as
\begin{equation}
	\begin{split}
		B=-W\frac{\partial p}{\partial W}
		=W\frac{\partial^2 E_{\rm TF}}{\partial W^2}.
	\end{split}
\end{equation}
Substituting $E_{\rm TF}(W)$ into this definition gives
\begin{equation}
	\begin{split}
		B(W)
		=g\frac{N^2}{W^2}
		-\frac{1}{2}\frac{N^{3/2}}{W^{3/2}} .
	\end{split}
\end{equation}
Finally, by evaluating this expression at the equilibrium length $W_e=9Ng^2/4$, we obtain
\begin{equation}
	\begin{split}
		B_{\text{TF}}
		=\frac{4}{81g^3}.
		\label{Blim}
	\end{split}
\end{equation}

\section{Validation of the spatial-scaling extraction of the modulus \(B\)}
\label{SSM-B}

In this appendix, we show that the spatial-scaling method used to extract the modulus does not change the definition of \(B\) near the equilibrium state. 
We introduce the scaled coordinate and wave function as
\begin{equation}
	\begin{split}
		x'=ax,\qquad 
		\psi'(x')=\frac{1}{\sqrt{a}}\psi(x),
	\end{split}
\end{equation}
which preserves the norm of the wave function. 
The energy of the scaled state can be written as
\begin{equation}
	\begin{split}
		E'(a)
		=&\int \left[
		\frac{1}{2}\left|\frac{\partial \psi'}{\partial x'}\right|^2
		+\frac{1}{2}g|\psi'|^4
		-\frac{2}{3}|\psi'|^3
		\right]dx'  \\
		=&\int \left[
		\frac{1}{2a^2}|\partial_x\psi|^2
		+\frac{1}{2a}g|\psi|^4
		-\frac{2}{3\sqrt{a}}|\psi|^3
		\right]dx .
	\end{split}
\end{equation}
Meanwhile, using the definition of the effective length, one obtains 
\begin{equation} 
	W'(a) =\frac{\left(\int |\psi'(x')|^2 dx'\right)^2} {\int |\psi'(x')|^4 dx'} =\frac{\left(\int |\psi(x)|^2 dx\right)^2} {a^{-1}\int |\psi(x)|^4 dx} =aW . 
\end{equation}
Therefore,
\begin{equation}
	\begin{split}
		\frac{\partial E'}{\partial W'}
		=\frac{\partial E'/\partial a}{\partial W'/\partial a}
		=\frac{1}{W}\frac{\partial E'}{\partial a},
		\frac{\partial^2 E'}{\partial W'^2}
		=\frac{1}{W^2}\frac{\partial^2 E'}{\partial a^2}.
	\end{split}
\end{equation}
The modulus extracted from the scaled energy curve is thus
\begin{equation}
	B'
	=\left(W'\frac{\partial^2 E'}{\partial W'^2}\right)_{a=1}
	=\left(\frac{aW}{W^2}\frac{\partial^2 E'}{\partial a^2}\right)_{a=1}
	=\frac{1}{W}\left(\frac{\partial^2 E'}{\partial a^2}\right)_{a=1}.
\end{equation}
At \(a=1\), the scaled state reduces to the original equilibrium state, so that \(W'=W\) and \(E'(1)=E\). 
Moreover, the scaled energy \(E'(a)\) represents the energy of the same state along the effective-length direction \(W'=aW\). 
Therefore,
\begin{equation}
	\begin{split}
		\left(\frac{\partial^2 E}{\partial W^2}\right)_{W=W_e}
		=
		\left(\frac{\partial^2 E'}{\partial W'^2}\right)_{a=1}
		=
		\frac{1}{W^2}
		\left(\frac{\partial^2 E'}{\partial a^2}\right)_{a=1}.
	\end{split}
\end{equation}
Using the definition
\begin{equation}
	B=W\left(\frac{\partial^2 E}{\partial W^2}\right)_{W=W_e},
\end{equation}
we obtain
\begin{equation}
	B
	=W
	\left(\frac{\partial^2 E'}{\partial W'^2}\right)_{a=1}
	=\frac{1}{W}
	\left(\frac{\partial^2 E'}{\partial a^2}\right)_{a=1}
	=B' .
\end{equation}
Thus, the modulus obtained from the scaled energy curve is identical to the modulus of the original equilibrium state.

This equivalence shows that, by sampling several scaled states around \(a=1\), one obtains the energy curve \(E(W)\), from which the curvature \(\partial^2 E/\partial W^2\) can be extracted by a quadratic fit. 
The bulk modulus is then calculated as
\begin{equation}
	B=W\frac{\partial^2 E}{\partial W^2}.
\end{equation}
This procedure is referred to as the spatial scaling method for extracting the modulus \(B\) in the main text.

\section{Derivation of the effective-width expression \(W_{N,g}\)}
\label{WNg-derivation}

In this appendix, we derive an analytical approximation for the equilibrium effective width as a function of the particle number \(N\) and the interaction strength \(g\). Within the VA framework, the 
equilibrium width is determined by the stationary condition \(\ddot{w}=0\), which gives 
\begin{equation} 
	C_2\alpha+C_3gNw+C_4N^{1/2}w^{3/2}=0 . \label{key} 
\end{equation} 
Although the TF approximation gives the leading-order result \(W_{\rm TF}=9Ng^2/4\), it shows noticeable deviations from the numerical width, especially in the small-\(N\) regime. We therefore introduce a corrected 
effective-width expression in the form 
\begin{equation} 
	W_{N,g}=\frac{9}{4}Ng^2+f(N,g), \label{WNg_ansatz} 
\end{equation} 
where \(f(N,g)\) accounts for the correction beyond the pure TF limit. To determine the scaling of this correction, we consider the characteristic point associated with the crossover from the 
soliton-like regime to the droplet regime. At this point, the equilibrium 
width reaches a local minimum as a function of \(N\) for a given \(g\), and thus satisfies \(\partial w/\partial N=0\). Taking the derivative of Eq.~(\ref{key}) with respect to \(N\) at fixed \(g\), and 
imposing \(\partial w/\partial N=0\), we obtain 
\begin{equation} 
	C_3gNw+\frac{C_4}{2}N^{1/2}w^{3/2}=0 . 
\end{equation}
Together with Eq.~(\ref{key}), this condition yields the characteristic scaling relation
\begin{equation}
	\begin{split}
		w
		=
		\left(-\frac{2C_2\alpha}{C_4}\right)^{2/3}N^{-1/3}
		=
		\left(\frac{2C_3}{C_4}\right)^2g^2N .
		\label{wN_scaling}
	\end{split}
\end{equation}
Equivalently, along the characteristic crossover curve in the \((N,g)\) parameter plane, one has
\begin{equation}
	\begin{split}
		g^2
		=
		\left(
		\frac{C_2\alpha C_4^2}{4C_3^3N^2}
		\right)^{2/3}.
		\label{gN_scaling}
	\end{split}
\end{equation}
Here, Eq.~(\ref{gN_scaling}) should be understood as a characteristic scaling relation on the crossover curve, rather than a constraint on the independent control parameters \(N\) and \(g\).

We now use this scaling relation to determine the correction term in Eq.~(\ref{WNg_ansatz}). 
Since the effective width also has a local minimum at the same characteristic point, we impose
\begin{equation}
	\begin{split}
		\left(\frac{\partial W_{N,g}}{\partial N}\right)_g=0 .
	\end{split}
\end{equation}
Substituting Eq.~(\ref{WNg_ansatz}) into this condition gives
\begin{equation}
	\begin{split}
		\frac{9}{4}g^2+\frac{\partial f}{\partial N}=0 .
	\end{split}
\end{equation}
Using Eq.~(\ref{gN_scaling}), we obtain
\begin{equation}
	\begin{split}
		\frac{\partial f}{\partial N}
		=
		-\frac{9}{4}
		\left(
		\frac{C_2\alpha C_4^2}{4C_3^3N^2}
		\right)^{2/3}.
	\end{split}
\end{equation}
Integrating this expression with respect to \(N\), one obtains
\begin{equation}
	\begin{split}
		f(N,g)
		=
		\frac{27}{4}
		\left(
		\frac{C_2\alpha C_4^2}{4C_3^3N^{1/2}}
		\right)^{2/3}
		+C(g),
	\end{split}
\end{equation}
where \(C(g)\) is an integration function depending only on \(g\). 
In the present work, we take \(C(g)=g/2\), which leads to
\begin{equation}
	\begin{split}
		W_{N,g}=\frac{9}{4}Ng^2+\frac{27}{4}\left(\frac{C_2\alpha C_4^2}{4C_3^3N^{1/2}}\right)^{2/3}+\frac{1}{2}g.
		\label{W_VA_TF}
	\end{split}
\end{equation}

\section*{CRediT authorship contribution statement}

\textbf{Rui Zhang}: Writing -- original draft, Investigation, Data curation, Software, Visualization.
\textbf{Tianmiao Zhang}: Writing -- original draft, Investigation, Data curation, Formal analysis, Validation.
\textbf{Huanbo Luo}: Methodology, Supervision, Writing -- review \& editing, Funding acquisition.
\textbf{Zibin Zhao}: Conceptualization, Methodology, Formal analysis, Writing -- review \& editing.

\section*{Declaration of competing interest}

The authors declare that they have no known competing financial interests or personal relationships that could have appeared to influence the work reported in this paper.

\section*{Acknowledgments}
This work was
supported by NNSFC (China) through Grants No. 12274077, No. 12475014,
Guangdong Basic and Applied Basic Research Foundation No. 2024A1515030131,
No. 2025A1515011128, No. 2023A1515110198, No. 2023A1515010770, the Research Fund of
Guangdong-Hong Kong-Macao Joint Laboratory for Intelligent Micro-Nano
Optoelectronic Technology through grant No. 2020B1212030010. 	
	
\section*{Data availability}

Data will be made available on request.


\begin{thebibliography}{99}
	\bibitem{Elasticity} Landau LD, Lifshitz EM, Kosevich AM, Pitaevskii LP. Theory of Elasticity. 3rd ed. Course of Theoretical Physics, vol. 7. Oxford: Butterworth-Heinemann; 1986. Chapter I, pp. 1--37. https://doi.org/10.1016/C2009-0-25521-8.
	\bibitem{Continuum}Reddy JN. An Introduction to Continuum Mechanics. 2nd ed. Cambridge: Cambridge University Press; 2013. https://doi.org/10.1017/CBO9781139178952.
	\bibitem{LHY-3d} Petrov DS. Quantum Mechanical Stabilization of a Collapsing Bose-Bose Mixture. Phys Rev Lett 2015;115:155302. https://doi.org/10.1103/PhysRevLett.115.155302.
	\bibitem{LHY-1D} Petrov DS, Astrakharchik GE. Ultradilute Low-Dimensional Liquids. Phys Rev Lett 2016;117:100401. https://doi.org/10.1103/PhysRevLett.117.100401.
	\bibitem{Ferrier2016} Ferrier-Barbut I, Kadau H, Schmitt M, Wenzel M, Pfau T. Observation of quantum droplets in a strongly dipolar Bose gas. Phys Rev Lett. 2016;116:215301. https://doi.org/10.1103/PhysRevLett.116.215301.
	\bibitem{binary-QD} Cabrera CR, Tanzi L, Sanz J, Naylor B, Thomas P, Cheiney P, et al. Quantum liquid droplets in a mixture of Bose-Einstein condensates. Science 2018;359:301-4. https://doi.org/10.1126/science.aao5686.
	\bibitem{QD-PRL}Semeghini G, Ferioli G, Masi L, Mazzinghi C, Wolswijk L, Minardi F, et al. Self-Bound Quantum Droplets of Atomic Mixtures in Free Space. Phys Rev Lett 2018;120:235301. https://doi.org/10.1103/PhysRevLett.120.235301.
	\bibitem{Cavicchioli2025} Cavicchioli L, Fort C, Ancilotto F, Modugno M, Minardi F, Burchianti A. Dynamical Formation of Multiple Quantum Droplets in a Bose-Bose Mixture. Phys Rev Lett 2025;134:093401. https://doi.org/10.1103/PhysRevLett.134.093401.
	\bibitem{Ferioli2020}Ferioli G, Semeghini G, Terradas-Briansó S, Masi L, Fattori M, Modugno M. Dynamical formation of quantum droplets in a $^{39}\mathrm{K}$ mixture. Phys Rev Res 2020;2:013269. https://doi.org/10.1103/PhysRevResearch.2.013269.
	\bibitem{Ferioli2019} Ferioli G, Semeghini G, Masi L, Giusti G, Modugno G, Inguscio M, et al. Collisions of Self-Bound Quantum Droplets. Phys Rev Lett 2019;122:090401. https://doi.org/10.1103/PhysRevLett.122.090401.
	\bibitem{Dong2021} Dong L, Kartashov YV. Rotating Multidimensional Quantum Droplets. Phys Rev Lett 2021;126:244101. https://doi.org/10.1103/PhysRevLett.126.244101.
	\bibitem{Ma2023} Ma Y, Cui X. Quantum-fluctuation-driven dynamics of droplet splashing, recoiling, and deposition in ultracold binary Bose gases. Phys Rev Res 2023;5:013100. https://doi.org/10.1103/PhysRevResearch.5.013100.
	\bibitem{3D-Bulkmodulus} Zhao Z, Li G, Chen Z, Luo H-B, Liu B, Malomed BA, et al. Bulk modulus of three-dimensional quantum droplets. Phys Rev A 2026;113:043315. https://doi.org/10.1103/cbg5-9r8v.
	\bibitem{LZH-FOP} Luo Z-H, Pang W, Liu B, Li Y-Y, Malomed BA. A new form of liquid matter: Quantum droplets. Front Phys 2020;16:32201. https://doi.org/10.1007/s11467-020-1020-2.
	\bibitem{kartashov-3D} Kartashov YV, Malomed BA, Tarruell L, Torner L. Three-dimensional droplets of swirling superfluids. Phys Rev A 2018;98:013612. https://doi.org/10.1103/PhysRevA.98.013612.
	\bibitem{Dong-3D}Dong L, Fan M, Malomed BA. Three-dimensional vortex and multipole quantum droplets in a toroidal potential. Chaos, Solitons Fractals 2024;188:115499. https://doi.org/10.1016/j.chaos.2024.115499.
	\bibitem{Ma-shell} Ma Y, Cui X. Shell-Shaped Quantum Droplet in a Three-Component Ultracold Bose Gas. Phys Rev Lett 2025;134:043402. https://doi.org/10.1103/PhysRevLett.134.043402.
	\bibitem{Otajonov-3D} Otajonov SR. Quantum droplets in three-dimensional Bose-Einstein condensates. J Phys B: At Mol Opt Phys 2022;55:085001. https://doi.org/10.1088/1361-6455/ac6365.
	\bibitem{Guo-LHY} Guo Z, Jia F, Li L, Ma Y, Hutson JM, Cui X, et al. Lee-Huang-Yang effects in the ultracold mixture of $^{23}\mathrm{Na}$ and $^{87}\mathrm{Rb}$ with attractive interspecies interactions. Phys Rev Res 2021;3:033247. https://doi.org/10.1103/PhysRevResearch.3.033247.
	\bibitem{Astrakharchik-dynamical} Astrakharchik GE, Malomed BA. Dynamics of one-dimensional quantum droplets. Phys Rev A 2018;98:013631. https://doi.org/10.1103/PhysRevA.98.013631.
	\bibitem{Parisi-QDs-1D} Parisi L, Giorgini S. Quantum droplets in one-dimensional Bose mixtures: A quantum Monte Carlo study. Phys Rev A 2020;102:023318. https://doi.org/10.1103/PhysRevA.102.023318.
	\bibitem{Mistakidis-formation} Mistakidis SI, Mithun T, Kevrekidis PG, Sadeghpour HR, Schmelcher P. Formation and quench of homonuclear and heteronuclear quantum droplets in one dimension. Phys Rev Res 2021;3:043128. https://doi.org/10.1103/PhysRevResearch.3.043128.
	\bibitem{Tylutki-collective} Tylutki M, Astrakharchik GE, Malomed BA, Petrov DS. Collective excitations of a one-dimensional quantum droplet. Phys Rev A 2020;101:051601. https://doi.org/10.1103/PhysRevA.101.051601.
	\bibitem{Pylak-crossover} Pylak M, Gajda M, Zin P. Dipolar droplets at the crossover from three dimensions to one dimension. Phys Rev A 2024;110:063322. https://doi.org/10.1103/PhysRevA.110.063322.
	\bibitem{Ilg-crossover} Ilg T, Kumlin J, Santos L, Petrov DS, Büchler HP. Dimensional crossover for the beyond-mean-field correction in Bose gases. Phys Rev A 2018;98:051604. https://doi.org/10.1103/PhysRevA.98.051604.
	\bibitem{Lin-2d} Lin Z, Xu X, Chen Z, Yan Z, Mai Z, Liu B. Two-dimensional vortex quantum droplets get thick. Commun Nonlinear Sci Numer Simul 2021;93:105536. https://doi.org/10.1016/j.cnsns.2020.105536.
	\bibitem{Li-2dvortex} Li Y, Chen Z, Luo Z, Huang C, Tan H, Pang W, et al. Two-dimensional vortex quantum droplets. Phys Rev A 2018;98:063602. https://doi.org/10.1103/PhysRevA.98.063602.
	\bibitem{Dong-Internal} Dong L, Shi K, Huang C. Internal modes of two-dimensional quantum droplets. Phys Rev A 2022;106:053303. https://doi.org/10.1103/PhysRevA.106.053303.
	\bibitem{Hu-Collisional} Hu Y, Fei Y, Chen X-L, Zhang Y. Collisional dynamics of symmetric two-dimensional quantum droplets. Front Phys 2022;17:61505. https://doi.org/10.1007/s11467-022-1192-z.
	\bibitem{Stumer-2D} St\"urmer P, Tengstrand MN, Sachdeva R, Reimann SM. Breathing mode in two-dimensional binary self-bound Bose-gas droplets. Phys Rev A 2021;103:053302. https://doi.org/10.1103/PhysRevA.103.053302.
	\bibitem{Examilioti-2D} Examilioti P, Kavoulakis GM. Ground state and rotational properties of two-dimensional self-bound quantum droplets. J Phys B At Mol Opt Phys 2020;53. https://doi.org/10.1088/1361-6455/ab9766.
	\bibitem{Pitaevskii-BEC-book} Pitaevskii L, Stringari S. Bose-Einstein Condensation and Superfluidity. Oxford University Press; 2016. https://doi.org/10.1093/acprof:oso/9780198758884.001.0001.
	\bibitem{Boris-Multidimensional-soliton} Malomed BA. Multidimensional Solitons. AIP Publishing LLC; n.d. https://doi.org/10.1063/9780735425118.
	\bibitem{Mithun-Statistical} Mithun T, Mistakidis SI, Schmelcher P, Kevrekidis PG. Statistical mechanics of one-dimensional quantum droplets. Phys Rev A 2021;104:033316. https://doi.org/10.1103/PhysRevA.104.033316.
	\bibitem{Otajonov-stationary} Otajonov SR, Tsoy EN, Abdullaev FKh. Stationary and dynamical properties of one-dimensional quantum droplets. Phys Lett A 2019;383:125980. https://doi.org/10.1016/j.physleta.2019.125980.
	\bibitem{Edmonds-QDs} Edmonds M, Bland T, Parker N. Quantum droplets of quasi-one-dimensional dipolar Bose-Einstein condensates. J Phys Commun 2020;4:125008. https://doi.org/10.1088/2399-6528/abcc3b.
	\bibitem{ZhaoF-Discrete} Zhao F, Yan Z, Cai X, Li C, Chen G, He H, et al. Discrete quantum droplets in one-dimensional optical lattices. Chaos, Solitons Fractals 2021;152:111313. https://doi.org/10.1016/j.chaos.2021.111313.
	\bibitem{Morera-universal} Morera I, Juli\'a-D\'iaz B, Valiente M. Universality of quantum liquids and droplets in one dimension. Phys Rev Res 2022;4:L042024. https://doi.org/10.1103/PhysRevResearch.4.L042024.
	\bibitem{Charalampidis-2component} Charalampidis EG, Mistakidis SI. Two-component droplet phases and their spectral stability in one dimension. Phys Rev A 2025;111. https://doi.org/10.1103/PhysRevA.111.013318.
	\bibitem{Katsimiga-interaction} Katsimiga GC, Mistakidis SI, Malomed BA, Frantzeskakis DJ, Carretero-Gonzalez R, Kevrekidis PG. Interactions and Dynamics of One-Dimensional Droplets, Bubbles and Kinks. Condensed Matter 2023;8:67. https://doi.org/10.3390/condmat8030067.
	\bibitem{Lv-Breather} Lv L-Z, Gao P, Yang Z-Y, Yang W-L. Breather excitations on the one-dimensional quantum droplet. Phys Lett A 2022;438:128124. https://doi.org/10.1016/j.physleta.2022.128124.
	\bibitem{Depalo-quasi-1D} De Palo S, Orignac E, Citro R. Formation and fragmentation of quantum droplets in a quasi-one-dimensional dipolar Bose gas. Phys Rev B 2022;106:014503. https://doi.org/10.1103/PhysRevB.106.014503.
	\bibitem{DuX-ground} Du X, Fei Y, Chen X-L, Zhang Y. Ground-state properties and Bogoliubov modes of a harmonically trapped one-dimensional quantum droplet. Phys Rev A 2023;108:033312. https://doi.org/10.1103/PhysRevA.108.033312.
	\bibitem{BEC-bulk} Pethick CJ, Smith H. Bose-Einstein Condensation in Dilute Gases. 2nd ed. Cambridge: Cambridge University Press; 2008. https://doi.org/10.1017/CBO9780511802850.
	\bibitem{ft-bec} Baizakov BB, Bouketir A, Messikh A, Benseghir A, Pumarov BA. Variational analysis of flat-top solitons in Bose-Einstein condensates. Int J Mod Phys B 2011;25:2427-2440. https://doi.org/10.1142/S0217979211101521.
	\bibitem{ITP} Yang J. Nonlinear Waves in Integrable and Nonintegrable Systems. Society for Industrial and Applied Mathematics; 2010. https://doi.org/10.1137/1.9780898719680.
	\bibitem{Rao-mechanical} Rao S. Mechanical Vibrations in SI Units. Pearson Deutschland; 2017.
	\bibitem{Sinha-quasi-1D} Sinha S, Santos L. Cold Dipolar Gases in Quasi-One-Dimensional Geometries. Phys Rev Lett 2007;99:140406. https://doi.org/10.1103/PhysRevLett.99.140406.
	\bibitem{LiG-anisotropic} Li G, Zhao Z, Jiang X, Chen Z, Liu B, Malomed BA, et al. Strongly Anisotropic Vortices in Dipolar Quantum Droplets. Phys Rev Lett 2024;133:053804. https://doi.org/10.1103/PhysRevLett.133.053804.
	\bibitem{Li-2D} Li G, Jiang X, Liu B, Chen Z, Malomed BA, Li Y. Two-dimensional anisotropic vortex quantum droplets in dipolar Bose-Einstein condensates. Front Phys 2024;19:22202. https://doi.org/10.1007/s11467-023-1338-7.
	\bibitem{CaiY} Cai Y, Yuan Y, Rosenkranz M, Pu H, Bao W. Vortex patterns and the critical rotational frequency in rotating dipolar Bose-Einstein condensates. Phys Rev A 2018;98:023610. https://doi.org/10.1103/PhysRevA.98.023610.
	\bibitem{Ground-state} W\"achtler F, Santos L. Ground-state properties and elementary excitations of quantum droplets in dipolar Bose-Einstein condensates. Phys Rev A 2016;94:043618. https://doi.org/10.1103/PhysRevA.94.043618.
	\bibitem{Bisset-Ground}Bisset RN, Wilson RM, Baillie D, Blakie PB. Ground-state phase diagram of a dipolar condensate with quantum fluctuations. Phys Rev A 2016;94:033619. https://doi.org/10.1103/PhysRevA.94.033619.
	\bibitem{Petter-Probing} Petter D, Natale G, van Bijnen RMW, Patscheider A, Mark MJ, Chomaz L, et al. Probing the Roton Excitation Spectrum of a Stable Dipolar Bose Gas. Phys Rev Lett 2019;122. https://doi.org/10.1103/PhysRevLett.122.183401.
	\bibitem{DinizPC} Diniz PC, Oliveira EAB, Lima ARP, Henn EAL. Ground state and collective excitations of a dipolar Bose-Einstein condensate in a bubble trap. Sci Rep 2020;10. https://doi.org/10.1038/s41598-020-61657-0.
	\bibitem{Baillie-collective} Baillie D, Wilson RM, Blakie PB. Collective Excitations of Self-Bound Droplets of a Dipolar Quantum Fluid. Phys Rev Lett 2017;119. https://doi.org/10.1103/PhysRevLett.119.255302.
	\bibitem{ZjangF} Zhang F, Yin L. Phonon Stability of Quantum Droplets in Dipolar Bose Gases. Chin Phys Lett 2022;39:060301. https://doi.org/10.1088/0256-307X/39/6/060301.
	\bibitem{Ferrier-Scissors} Ferrier-Barbut I, Wenzel M, Böttcher F, Langen T, Isoard M, Stringari S, et al. Scissors Mode of Dipolar Quantum Droplets of Dysprosium Atoms. Phys Rev Lett 2018;120:160402. https://doi.org/10.1103/PhysRevLett.120.160402.
	\bibitem{Blakie-axial} Blakie PB. Axial Collective Mode of a Dipolar Quantum Droplet. Photonics 2023;10:393. https://doi.org/10.3390/photonics10040393.
	\bibitem{Jaksch-OL} Jaksch D, Bruder C, Cirac JI, Gardiner CW, Zoller P. Cold Bosonic Atoms in Optical Lattices. Phys Rev Lett 1998;81:3108-11. https://doi.org/10.1103/PhysRevLett.81.3108.
	\bibitem{Edwards-VA-ol} Edwards M, DeBeer LM, Demenikov M, Galbreath J, Mahaney TJ, Nelsen B, et al. A hybrid Lagrangian variational method for Bose-Einstein condensates in optical lattices. J Phys B: At Mol Opt Phys 2005;38:363. https://doi.org/10.1088/0953-4075/38/4/004.
	\bibitem{Nie-OL} Nie Y, Zheng J-H, Yang T. Spectra and dynamics of quantum droplets in an optical lattice. Phys Rev A 2023;108. https://doi.org/10.1103/PhysRevA.108.053310.
	\bibitem{QDs-1d-OL} Vall\'es-Muns J, Morera I, Astrakharchik GE, Juli\'a-D\'iaz B. Quantum droplets with particle imbalance in one-dimensional optical lattices. SciPost Physics 2024;16:074. https://doi.org/10.21468/SciPostPhys.16.3.074.
	\bibitem{ZhouZ-dynamical-ol} Zhou Z, Yu X, Zou Y, Zhong H. Dynamics of quantum droplets in a one-dimensional optical lattice. Commun Nonlinear Sci Numer Simul 2019;78:104881. https://doi.org/10.1016/j.cnsns.2019.104881.
	\bibitem{ZhouZ-Controllable} Zhou Z, Shi Y, Tang S, Deng H, Wang H, He X, et al. Controllable dissipative quantum droplets in one-dimensional optical lattices. Chaos, Solitons Fractals 2021;150:111193. https://doi.org/10.1016/j.chaos.2021.111193.
	\bibitem{Ancilotto-vortex-ol} Ancilotto F. Controlling quantum vortex dynamics and vortex-antivortex annihilation in Bose-Einstein condensates with optical lattices. Phys Rev A 2024;110. https://doi.org/10.1103/PhysRevA.110.013302.
	\end{thebibliography}
\end{document}